\documentclass[12pt,epsf,fleqn]{article}
\usepackage{epsfig}
\usepackage[dvips]{color}
\begin{document}
\title{Electroweak phase transition in the MNMSSM with explicit CP violation}
\author{S.W. Ham$^{(1)}$, J.O. Im$^{(2)}$, and S.K. Oh$^{(1,2)}$
\\
\\
{\it $^{\rm (1)}$ Center for High Energy Physics, Kyungpook National University}\\
{\it Daegu 702-701, Korea} \\
{\it $^{\rm (2)}$ Department of Physics, Konkuk University, Seoul 143-701, Korea}
\\
\\
}
\date{}
\maketitle
\begin{abstract}
In explicit CP violation scenario of the minimal non-minimal supersymmetric standard model (MNMSSM),
the possibility of a strongly first-order electroweak phase transition (EWPT) is investigated
at the one-loop level, where the radiative corrections from the loops of the top and stop quarks
are taken into account.
Assuming that the stop quark masses are not degenerate, the radiative corrections
due to the stop quarks give rise to a CP phase, which triggers the scalar-pseudoscalar mixing
in the Higgs sector of the MNMSSM.
The lighter stop quark need not always to have a small mass in order to ensure the strongly first-order EWPT.
In the MNMSSM with explicit CP violation, it is found that the strength of the first-order EWPT depends
on several factors, such as the lightest neutral Higgs boson mass and the nontrivial CP phase arising
from stop quark masses. The effects of these factors are discussed.
\end{abstract}
\vfil
\eject

\section{Introduction}

For theoretical models to explain successfully the observed baryon asymmetry of the universe,
it is necessary that they should satisfy the conditions which have been suggested
by Sakharov several decades ago [1].
The three conditions for dynamically generating the baryon asymmetry are: the violation
of baryon number conservation, the violation of both C and CP,
and the deviation from thermal equilibrium.

The baryogenesis via the electroweak phase transition (EWPT) [2-11],
which in principle may satisfy the Sakharov condition,
has been studied by many authors.
As is well known, in order to ensure sufficient deviation from thermal equilibrium,
the EWPT should be strongly first order, since otherwise the baryon asymmetry generated
during the electroweak phase transition would subsequently disappear.
In general, the strength of the EWPT is measured by comparing the vacuum expectation value (VEV)
of the Higgs field at the broken-phase state with the critical temperature.
The electroweak phase transition is said to be strong if the former is larger than the latter.

The Standard Model (SM) has already been investigated whether it can realize the strongly first-order EWPT.
It is found, however, that the SM faces severe difficulties to accommodate the desired EWPT,
because it cannot make the EWPT strong enough unless the mass of the SM Higgs boson is
below the present experimental lower bound.
Thus, in the SM, the EWPT is weakly first or higher order for the experimentally
allowed mass of the Higgs boson [12-17].
Also, the SM cannot produce the CP violation large enough to generate the baryon asymmetry
by means of the complex phase in the Cabibbo-Kobayashi-Maskawa matrix alone [18].

Many scenarios for expanding the SM have been studied in order to formulate
the idea of baryogenesis via the EWPT [19-33].
Embracing the supersymmetry (SUSY) is obviously the most plausible possibility to expand the SM.
As the simplest version of the supersymmetric SM,
the minimal supersymmetric standard model (MSSM) is studied widely
in the context of baryogenesis via the EWPT.
In the MSSM, a strongly first-order EWPT is only possible in a confined parameter region
where a stop quark is lighter than the top quark and the other stop quark is
as heavy as the SUSY breaking scale (1-2 TeV).

The possibility of the EWPT is further examined in other versions of the supersymmetric SM,
by introducing one or several Higgs singlets to the Higgs sector of the MSSM,
such as the next-to-minimal supersymmetric SM (NMSSM) with a $Z_3$-symmetry [24-27],
the general NMSSM with  broken $Z_3$ [28, 29], the minimal non-minimal supersymmetric SM (MNMSSM or nMSSM)
with tadpole terms [30, 31], and the minimal supersymmetric models with an additional $U(1)'$ [32],
to cite some of them.
These models are different from the MSSM in the sense that they do not always need a light stop quark
in order to produce a strongly first-order EWPT.
We also have studied the MNMSSM with a tadpole term elsewhere for the possibility of the EWPT [31],
with no CP mixing in its Higgs sector.
We have found that this model has some parameter regions where a sufficiently strong first-order EWPT
may occur for electroweak baryogenesis, without requiring a light stop quark.

In this paper, we would like to continue the previous study so as to consider the case
with explicit CP violation in the Higgs sector of the MNMSSM at the one-loop level.
It is known that there is no CP mixing between the scalar and pseudoscalar Higgs fields
in the MNMSSM at the tree level.
At the one-loop level, the MNMSSM may have the CP violation in an explicit way, generated
by the radiative contributions due to the top and stop quark loops.
In this explicit CP violation scenario, we examine if it is possible
to achieve the EWPT that is first order and strong enough to realize the necessary electroweak baryogenesis.

We find that a strongly first-order EWPT is indeed possible in the MNMSSM
with explicit CP violation at the one-loop level.
The trilinear term in the tree-level Higgs potential plays an important role,
later at the one-loop level, in achieving a strongly first order EWPT
in the MNMSSM with explicit CP violation.
The nontrivial CP phase arising from the stop quark masses alters
the strength of a first-order EWPT to be even stronger a little.
The strength of the first-order EWPT depends also on the lightest neutral Higgs boson mass
and the stop quark masses, but not in a definite way.
As the lightest Higgs boson mass increases,
the strength of the first-order EWPT may either increase or decrease.
Also, as the two stop quark masses increase, it may either increase or decrease.

\section{THE CP VIOLATING HIGGS POTENTIAL}

The Higgs sector of the MNMSSM consists of two Higgs doublet superfields $H_1 = (H_1^0, H^-)$,
$H_2 = (H^+, H_2^0)$ and a Higgs singlet superfield $N$ [34-36].
Keeping only the Yukawa couplings for the third generation of quarks, the superpotential of MNMSSM
may be written as
\begin{equation}
W = h_t Q H_2 t_R^c + h_b Q H_1 b_R^c + \lambda N H_1^T \epsilon H_2 \ ,
\end{equation}
where $h_q (q=t,b)$ are the quark Yukawa couplings, $\lambda$ is a dimensionless coupling constant,
$\epsilon$ is the antisymmetric $2 \times 2$ matrix with $\epsilon_{12} = 1$.
The left-handed quark superfields $q_L (q=t,b)$ form the SU(2) doublet $Q^T = (t_L, b_L)$,
and the charge conjugate of the right-handed $q_R (q=t,b)$ quark superfields form the SU(2) singlets $t_R^c$
and $b_R^c$.

The tree-level Higgs potential, $V_0$, of the MNMSSM may be decomposed into $D$-terms, $F$-terms,
the soft terms, and the tadpole terms as
\[
    V_0 = V_D + V_F + V_{\rm S} + V_{\rm T} \ ,
\]
where
\begin{eqnarray}
V_D & = & {g_2^2 \over 8} (H_1^{\dag} \vec \sigma H_1 + H_2^{\dag} \vec \sigma H_2)^2
+ {g_1^2\over 8}(|H_2|^2 - |H_1|^2)^2  \ , \cr
V_F & = & |\lambda|^2[(|H_1|^2+|H_2|^2)|N|^2+|H_1^T \epsilon H_2|^2] \ , \cr
V_{\rm S} & = & m_{H_1}^2|H_1|^2 + m_{H_2}^2|H_2|^2 + m_N^2|N|^2
- (\lambda A_{\lambda} H_1^T \epsilon H_2 N + {\rm H.c.} ) \ , \cr
V_{\rm T} & = &\mbox{} - (\xi^3 N + {\rm H.c.} )   \ ,
\end{eqnarray}
with $g_1$ and $g_2$ being the U(1) and SU(2) gauge coupling constants, respectively,
and $\vec \sigma$ being the Pauli matrices.
In $V_{\rm S}$, $A_{\lambda}$ is the trilinear soft SUSY breaking parameter with mass dimension,
and $m_{H_1}$, $m_{H_2}$, and $m_N$ are the soft SUSY breaking masses.
In $V_{\rm T}$, $\xi$ is the tadpole coefficient.

Without the tadpole terms, the above tree-level Higgs potential would additionally possess
a discrete symmetry which would then lead to a massless pseudo-Goldstone boson.
This is the unwanted axion, because the discrete symmetry is a Peccei-Quinn symmetry.
Generally, whenever a discrete symmetry is spontaneously broken down, an unwanted axion would appear.

In terms of the physical Higgs fields, the Higgs doublets and the Higgs singlet may be expressed as
\begin{eqnarray}
\begin{array}{lll}
        H_1 & = & \left ( \begin{array}{c}
          v_1 + S_1 + i \sin \beta P_1   \cr
          \sin \beta C^{+ *}
  \end{array} \right )  \ ,  \cr
        H_2 & = & \left ( \begin{array}{c}
          \cos \beta C^+           \cr
          (v_2 + S_2 + i \cos \beta P_1) e^{i \theta}
  \end{array} \right )   \ ,  \cr
        N & = & \left ( \begin{array}{c}
          x + S_3 + i P_2
  \end{array} \right ) e^{i \delta}   \ ,
\end{array}
\end{eqnarray}
where $S_i$ ($i$ = 1-3) are the scalar Higgs fields, $P_i$ ($i$=1, 2) are the pseudoscalar Higgs fields,
$C^+$ is the charged Higgs field, and $v_1$, $v_2$, and $x$ are respectively
the VEVs of $S_i$ ($i$ = 1-3) with $v = \sqrt{v_1^2 + v_2^2}$ = 174 GeV and $\tan \beta = v_2/v_1$.
Note that these VEVs are evaluated at zero temperature.
Thus, they will be denoted in the next section as $v_1(0)$, $v_2(0)$ and $x(0)$
to show explicitly the temperature dependence, whereas the VEVs at finite temperature
will be denoted as $v_1$, $v_2$, and $x$.

If complex phases are present in the Higgs sector of the MNMSSM,
we would expect CP violation in the MNMSSM.
The complex phases may explicitly appear from compex coefficients and/or
from the VEVs of the relevant Higgs fields.
In the tree-level Higgs potential, $A_{\lambda}$, $\lambda$, and $\xi$ can be complex.
Among them, it can be easily shown that the phase of $A_{\lambda}$ may be absorbed into
that of $\lambda$, without loss of generality.
Further, $\lambda$ can be made real by adjusting the phases of the Higgs doublets,
while $\xi$ can also be made real by adjusting the phase of the Higgs singlet.
Consequently, the Higgs potential of the MNMSSM at the tree level
can be made free of any CP violating complex phases.

The same conclusion can be drawn by applying the tree-level tadpole minimum conditions upon
the tree-level Higgs potential of the MNMSSM.
In general, five complex phases may arise from the tree-level Higgs potential of the MNMSSM
after spontaneous breakdown of the electroweak symmetry.
Besides the three complex coefficients $A_{\lambda} e^{i \phi_{A_{\lambda}}}$,
$\lambda e^{i \phi_{\lambda}}$, and $\xi e^{i \phi_{\xi}}$,
two of the VEVs can also be complex: $v_2 e^{i \theta}$, and $x e^{i \delta}$.

A pair of tree-level tadpole minimum conditions are derived from the first derivatives of
the tree-level Higgs potential with respect the two pseudoscalar Higgs fields as
\begin{eqnarray}
0 & = & 2 \lambda v x A_{\lambda} \sin \theta_1 \ , \cr
0 & = & \lambda v^2 A_{\lambda} \sin 2 \beta \sin \theta_1 + 2 \xi^3 \sin \theta_2 \ ,
\end{eqnarray}
where the two phases $\theta_1$ and $\theta_2$ are defined
as $\theta_1 = \phi_{\lambda} + \phi_{A_{\lambda}} + \theta + \delta$ and $\theta_2 = 3 \phi_{\xi} + \delta$.
It is straightforward that these two equations are satisfied only when both $\theta_1 = 0$ and $\theta_2 = 0$.

Therefore, either way, there is no complex phase that can induce CP mixing
between the scalar and pseudoscalar Higgs bosons in the the tree-level Higgs potential of the MNMSSM.
Thus, in order to accommodate any complex phases so as to induce the CP mixing
between the scalar and pseudoscalar Higgs bosons in the present model,
one has to consider higher-order corrections.
The radiative corrections due to the top and stop quarks is known
to affect significantly the tree-level Higgs sector of supersymmetric models.

The one-loop effective potential including the radiative corrections due to top and stop quarks is given by [37]
\begin{equation}
V^1 = \sum_{i = 1}^2 {3 {\cal M}_{{\tilde t}_i}^4 \over 32 \pi^2}
  \left (\log {{\cal M}_{{\tilde t}_i}^2 \over \Lambda^2} - {3\over 2} \right )
  - {3 {\cal M}_t^4 \over 16 \pi^2} \left (\log {{\cal M}_t^2 \over \Lambda^2}
  - {3\over 2} \right ) \ ,
\end{equation}
where ${\tilde t}_i$ ($i$ = 1 ,2) are stop quarks, $\Lambda$ is the renormalization scale
in the modified minimal subtraction scheme, and ${\cal M}_t$ and ${\cal M}_{{\tilde t}_i}$ respectively
are the top and stop quark masses given as functions of the Higgs fields.

After the spontaneous breakdown of the electroweak symmetry, the top quark mass is given
by $m_t = h_t v_2$ and the stop quark masses are given by
\begin{equation}
m_{{\tilde t}_1, {\tilde t}_2}^2 = m_T^2 + m_t^2 \mp h_t
\sqrt{A_t^2 v_2^2 + \lambda^2 v_1^2 x^2 + 2 \lambda A_t v_1 v_2 x \cos \phi_t}
\end{equation}
where $m_T$ is the soft SUSY breaking mass and $A_t$ is the trilinear SUSY breaking parameter
with mass dimension.

One may notice that a complex phase $\phi_t$ is already present in the stop quark masses.
It is given by $\phi_t = \phi_{A_t} + \phi_{\lambda} + \theta + \delta$,
where each phase is respectively originated from $A_t e^{i \phi_{A_t}}$,
$\lambda e^{i \phi_{\lambda}}$, $v_2 e^{i \theta}$, and $x e^{i \delta}$.

The CP-odd tadpole minimum conditions with respect to the pseudoscalar Higgs fields
are modified at the one-loop level as
\begin{eqnarray}
0 & = & A_{\lambda} \sin \theta_1 + {3 h_t^2 \over 16 \pi^2 } A_t \sin \phi_t
f(m_{{\tilde t}_1}^2, m_{{\tilde t}_2}^2) \ , \cr
0 & = & A_{\lambda} \lambda v^2 \sin 2 \beta \sin \theta_1 + 2 \xi^3 \sin \theta_2
+ {3 h_t^2 \over 16 \pi^2 } A_t \lambda v^2 \sin 2 \beta \sin \phi_t
f(m_{{\tilde t}_1}^2, m_{{\tilde t}_2}^2) \ ,
\end{eqnarray}
where the CP violating phase $\theta_1$, $\theta_2$, and $\phi_t$ are the same as in the tree level,
and the scale-dependent function $f(m_x^2, m_y^2)$ is defined by
\begin{equation}
 f(m_x^2, m_y^2) = {1 \over (m_y^2 - m_x^2)} \left[  m_x^2 \log {m_x^2 \over \Lambda^2} - m_y^2
\log {m_y^2 \over \Lambda^2} \right] + 1 \ .
\end{equation}

Substituting the first tadpole minimum condition into the second one,
one can easily see that the second tadpole minimum condition is satisfied by $\sin\theta_2 = 0$.
Further, the first tadpole minimum condition alone shows that $\theta_1$
at the one-loop level is not zero but dependent on the other parameters.
Consequently, the MNMSSM may eventually have one physical CP phase $\phi_t$
at the one-loop level even if it is free of any complex phase at the tree level.
We will rename $\theta_1$ at the one-loop level as $\phi_0$ hereafter,
in order to avoid any confusion with the tree-level $\theta_1$ that is zero.

Now, let us calculate the masses of the five neutral Higgs bosons.
They are given as the eigenvalues of the symmetric mass matrix $M$ that is calculated from
the second derivatives with respect to $S_1$, $S_2$, $S_3$, $P_1$ and $P_2$.
In the basis of ($S_1, S_2, P_1, S_3, P_2$), the matrix elements of $M$ are obtained as
\begin{eqnarray}
M_{11} & = & M_{11}^t + (m_Z \cos \beta)^2 + m_A^2 \sin^2 \beta , \cr
M_{22} & = & M_{22}^t + (m_Z \sin \beta)^2 + m_A^2 \cos^2 \beta , \cr
M_{33} & = & M_{33}^t + m_A^2  \ , \cr
M_{44} & = & M_{44}^t + {v^2 \over 4 x^2} m_A^2 \sin^2 2 \beta  + {\xi^3 \over x} \ , \cr
M_{55} & = & M_{55}^t + {v^2 \over 4 x^2} m_A^2 \sin^2 2 \beta  + {\xi^3 \over x} \ , \cr
M_{12} & = & M_{12}^t + (2 \lambda^2 v^2 - m_Z^2 - m_A^2) \sin \beta \cos \beta \ , \cr
M_{13} & = & M_{13}^t  \ , \cr
M_{14} & = & M_{14}^t - {v \over x} m_A^2 \sin^2 \beta \cos \beta + 2 v \lambda^2 x \cos \beta \ , \cr
M_{15} & = & M_{15}^t  \ , \cr
M_{23} & = & M_{23}^t  \ , \cr
M_{24} & = & M_{24}^t - {v \over x} m_A^2 \sin \beta \cos^2 \beta + 2 v \lambda^2 x \sin \beta \ , \cr
M_{25} & = & M_{25}^t  \ , \cr
M_{34} & = & M_{34}^t  \ , \cr
M_{35} & = & M_{35}^t + {v \over x} m_A^2 \sin \beta \cos \beta  \ , \cr
M_{45} & = & M_{45}^t  \ ,
\end{eqnarray}
where $m_Z^2 = (g_1^2 + g_2^2) v^2/2$ is the squared mass of the neutral gauge boson,
$m_A^2$ is introduced for convenience as
\begin{eqnarray}
m_A^2 & = & {\lambda x A_{\lambda} \cos \phi_0 \over \sin \beta \cos \beta}
+ {3 m_t^2 A_t \lambda x \cos \phi_t \over 16 \pi^2 v^2 \sin^3 \beta \cos \beta}
f (m_{{\tilde t}_1}^2, m_{{\tilde t}_2}^2)  \ ,
\end{eqnarray}
and $M^t_{ij}$ are the radiative corrections due to the top and stop quarks at the one-loop level.
Explicitly, $M^t_{ij}$ are given as
\begin{eqnarray}
M_{11}^t & = & {3 m_t^4 \lambda^2 x^2 \Delta_{{\tilde t}_1}^2 \over 8 \pi^2  v^2 \sin^2 \beta}
{g(m_{\tilde{t}_1}^2, \ m_{\tilde{t}_2}^2) \over (m_{\tilde{t}_2}^2 - m_{\tilde{t}_1}^2)^2}   \ , \cr
M_{22}^t & = & {3 m_t^4 A_t^2 \Delta_{{\tilde t}_2}^2 \over 8 \pi^2  v^2 \sin^2 \beta}
{g(m_{\tilde{t}_1}^2, \ m_{\tilde{t}_2}^2) \over (m_{\tilde{t}_2}^2 - m_{\tilde{t}_1}^2)^2}
+ {3 m_t^4 A_t \Delta_{{\tilde t}_2} \over 4 \pi^2 v^2 \sin^2 \beta}
{\log (m_{\tilde{t}_2}^2 / m_{\tilde{t}_1}^2)  \over (m_{\tilde{t}_2}^2 - m_{\tilde{t}_1}^2)} \cr
& &\mbox{} + {3 m_t^4 \over 8 \pi^2 v^2 \sin^2 \beta}
\log \left ( {m_{\tilde{t}_1}^2  m_{\tilde{t}_2}^2 \over m_t^4} \right ) \ , \cr
M_{33}^t & = & {3 m_t^4 \lambda^2 x^2 A_t^2 \sin^2 \phi_t \over 8 \pi^2 v^2 \sin^4 \beta}
{g(m_{\tilde{t}_1}^2, \ m_{\tilde{t}_2}^2) \over (m_{\tilde{t}_2}^2 - m_{\tilde{t}_1}^2 )^2}  \ , \cr
M_{44}^t & = & {3 m_t^4 \lambda^2 \Delta_{{\tilde t}_1}^2 \over 8 \pi^2 \tan^2 \beta}
{g(m_{\tilde{t}_1}^2, \ m_{\tilde{t}_2}^2) \over (m_{\tilde{t}_2}^2 - m_{\tilde{t}_1}^2 )^2}  \ , \cr
M_{55}^t & = & {3 m_t^4 \lambda^2 A_t^2 \sin^2 \phi_t \over 8 \pi^2 \tan^2 \beta}
{g(m_{\tilde{t}_1}^2, \ m_{\tilde{t}_2}^2) \over (m_{\tilde{t}_2}^2 - m_{\tilde{t}_1}^2 )^2}  \ ,  \cr
M_{12}^t & = & {3 m_t^4 \lambda x A_t \Delta_{{\tilde t}_1} \Delta_{{\tilde t}_2} \over 8 \pi^2 v^2 \sin^2 \beta}
{g(m_{\tilde{t}_1}^2, \ m_{\tilde{t}_2}^2) \over (m_{\tilde{t}_2}^2 - m_{\tilde{t}_1}^2)^2}
+ {3 m_t^4 \lambda x \Delta_{{\tilde t}_1} \over 8 \pi^2 v^2 \sin^2 \beta}
{\log (m_{\tilde{t}_2}^2 / m_{\tilde{t}_1}^2) \over (m_{\tilde{t}_2}^2 - m_{\tilde{t}_1}^2)}  \ , \cr
M_{13}^t & = & \mbox{} - {3 m_t^4 \lambda^2 x^2 A_t \Delta_{{\tilde t}_1} \sin \phi_t \over 8 \pi^2 v^2 \sin^3 \beta}
{g(m_{\tilde{t}_1}^2, \ m_{\tilde{t}_2}^2) \over (m_{\tilde{t}_2}^2 - m_{\tilde{t}_1}^2)^2 } \ , \cr
M_{14}^t & = & {3 m_t^4 \lambda^2 x \Delta_{{\tilde t}_1}^2 \over 8 \pi^2 v \sin \beta \tan \beta}
{g(m_{\tilde{t}_1}^2, \ m_{\tilde{t}_2}^2) \over (m_{\tilde{t}_2}^2 - m_{\tilde{t}_1}^2)^2 }
- {3 m_t^2 \lambda^2 x \cot \beta \over 8 \pi^2 v \sin \beta} f(m_{\tilde{t}_1}^2, \ m_{\tilde{t}_2}^2)  , \cr
M_{15}^t & = &\mbox{} - {3 m_t^4 \lambda^2 x A_t \Delta_{{\tilde t}_1} \sin \phi_t \over 8 \pi^2 v \sin \beta \tan \beta}
{g(m_{\tilde{t}_1}^2, \ m_{\tilde{t}_2}^2) \over (m_{\tilde{t}_2}^2 - m_{\tilde{t}_1}^2)^2 } \ , \cr
M_{23}^t & = & \mbox{} - {3 m_t^4 \lambda x A_t^2 \Delta_{{\tilde t}_2} \sin \phi_t \over 8 \pi^2 v^2 \sin^3 \beta}
{g(m_{\tilde{t}_1}^2, \ m_{\tilde{t}_2}^2) \over (m_{\tilde{t}_2}^2 - m_{\tilde{t}_1}^2)^2 }
- {3 m_t^4 \lambda x A_t \sin \phi _t \over 8 \pi^2 v^2 \sin^3 \beta}
{\log (m_{\tilde{t}_2}^2 / m_{\tilde{t}_1}^2) \over (m_{\tilde{t}_2}^2 - m_{\tilde{t}_1}^2)}  \ , \cr
M_{24}^t & = & {3 m_t^4 \lambda A_t \Delta_{{\tilde t}_1} \Delta_{{\tilde t}_2} \over 8 \pi^2 v \sin \beta \tan \beta}
{g(m_{\tilde{t}_1}^2, \ m_{\tilde{t}_2}^2) \over (m_{\tilde{t}_2}^2 - m_{\tilde{t}_1}^2)^2}
 + {3 m_t^4 \lambda \Delta_{{\tilde t}_1} \over 8 \pi^2 v \sin \beta \tan \beta}
{\log (m_{\tilde{t}_2}^2 / m_{\tilde{t}_1}^2) \over (m_{\tilde{t}_2}^2 - m_{\tilde{t}_1}^2) }  , \cr
M_{25}^t & = & \mbox{} - {3 m_t^4 \lambda A_t^2 \Delta_{{\tilde t}_2} \sin \phi_t \over 8 \pi^2 v \sin \beta \tan \beta}
{g(m_{\tilde{t}_1}^2, \ m_{\tilde{t}_2}^2) \over (m_{\tilde{t}_2}^2 - m_{\tilde{t}_1}^2)^2 }
- {3 m_t^4 \lambda A_t \sin \phi_t \over 8 \pi^2 v \sin \beta \tan \beta}
{ \log (m_{\tilde{t}_2}^2 / m_{\tilde{t}_1}^2) \over (m_{\tilde{t}_2}^2 - m_{\tilde{t}_1}^2)}  \ , \cr
M_{34}^t & = & \mbox{} - {3 m_t^4 \lambda^2 x A_t \Delta_{{\tilde t}_1} \sin \phi_t \over 8 \pi^2 v \sin^2 \beta \tan \beta}
{g(m_{\tilde{t}_1}^2, \ m_{\tilde{t}_2}^2) \over (m_{\tilde{t}_2}^2 - m_{\tilde{t}_1}^2)^2 } \ , \cr
M_{35}^t & = & \mbox{} {3 m_t^4 \lambda^2 x A_t^2 \sin^2 \phi_t \over 8 \pi^2 v \sin^2 \beta \tan \beta}
{g(m_{\tilde{t}_1}^2, \ m_{\tilde{t}_2}^2) \over (m_{\tilde{t}_2}^2 - m_{\tilde{t}_1}^2)^2 }  \ , \cr
M_{45}^t & = & \mbox{} - {3 m_t^4 \lambda^2 A_t \Delta_{{\tilde t}_1} \sin \phi_t \over 8 \pi^2 \tan^2 \beta}
{ g(m_{\tilde{t}_1}^2, \ m_{\tilde{t}_2}^2) \over (m_{\tilde{t}_2}^2 - m_{\tilde{t}_1}^2)^2 } \ ,
\end{eqnarray}
where
\begin{eqnarray}
 \Delta_{{\tilde t}_1} &=& A_t \cos \phi_t + \lambda x \cot \beta  \  , \cr
 \Delta_{{\tilde t}_2} & = & A_t + \lambda x \cot \beta \cos \phi_t \ ,
\end{eqnarray}
and $g(m_x^2,m_y^2)$ is another scale-independent function that is defined as
\begin{equation}
 g(m_x^2,m_y^2) = {m_y^2 + m_x^2 \over m_x^2 - m_y^2} \log {m_y^2 \over m_x^2} + 2 \ .
\end{equation}

The matrix elements of $M$ that are responsible for the CP mixing
between scalar and pseudoscalar Higgs fields in the ($S_1, S_2, P_1, S_3, P_2$) basis are
$M_{13}$, $M_{23}$, $M_{15}$, $M_{25}$, $M_{34}$, and $M_{45}$.
These matrix elements would be zero unless the radiative corrections $M^t$
due to the top and stop quarks at the one-loop level are taken into account.
Moreover, those matrix elements of $M^t$ for the CP mixing contain one CP phase $\phi_t$.
Thus, the magnitude of the scalar-pseudoscalar mixing is proportion to $\sin \phi_t$.
There would be no CP mixing between the scalar and pseudoscalar Higgs fields
in the Higgs sector of the present model in the cases of $\phi_t = 0$,
and the mixing would be maximal when $\sin \phi_t = 1$.

The five physical neutral Higgs bosons $h_i$ ($i$ = 1-5) would be defined
as the eigenstates of the mass matrix $M$.
The eigenvalues of $M$, denoted as $m^2_{h_i}$ ($i$ = 1-5), are their squared masses.
We sort these five neutral Higgs bosons in the increasing order of their masses
such that $m^2_{h_1}$ is the smallest eigenvalue and $h_1$ is the lightest neutral Higgs boson.
They are in general given as the mixtures of $S_1$, $S_2$, $S_3$, $P_1$, and $P_2$.
If $\phi_t = 0$, these five neutral Higgs bosons may be classified
into three scalar and two pseudoscalar Higgs bosons.
Unless $\phi_t$ vanishes, $h_i$ ($i$ = 1-5) would not have definite CP parities.

In order to measure the size of the CP violation, which arises from the CP mixing
between the scalar and pseudoscalar Higgs fields, it is preferable to define the following parameter,
\[
\rho = 5 (O_{11}^2 O_{12}^2 O_{13}^2 O_{14}^2 O_{15}^2)^{1/5} \ ,
\]
where $O_{ij}$ ($i,j$ =1-5) are the matrix elements of the orthogonal matrix $O$ that diagonalizes $M$.
The parameter $\rho$ can vary between 0 and 1, since $O_{ij}$ satisfy
the orthogonality condition of $\sum_{j = 1}^{5} O_{1j}^2$ = 1.
If $\rho = 0$, there would be no CP violation in the Higgs sector of the NMSSM.
On the other hand, if $\rho = 1$, CP symmetry would be maximally violated.
The maximal CP violation that leads to $\rho$ = 1 takes place
when $O_{11}^2 = O_{12}^2 = O_{13}^2 = O_{14}^2 = O_{15}^2 = 1/5$.

\section{THE THERMAL POTENTIAL}

Up to now, we have studied the Higgs sector of the MNMSSM at zero temperature.
The relevant quantities such as VEVs and various parameters are all defined and derived
at zero temperature.
In order to study the possibility of the EWPT in the present model,
we now consider the temperature dependence of the model.
In particular, the thermal effects due to the top and stop quarks are taken into account
for the thermal potential at finite temperature at the one-loop level.
We employ the effective potential method to obtain the radiatively corrected thermal potential
at the one-loop level.

At finite temperature $T$, the radiative corrections due to the top and stop quarks is
effectively given as [38]
\begin{equation}
V_T (\varphi, T)  =  \sum_{l = t, {\tilde t}_i} {n_l T^4 \over 2 \pi^2}
            \int_0^{\infty} dx \ x^2 \
            \log \left [1 \pm \exp{\left ( - \sqrt {x^2+{m_l^2(\varphi)/T^2 }} \right )  } \right ] \ ,
\end{equation}
where $\varphi$ denotes collectively the Higgs fields, the negative sign is
for the stop quarks ${\tilde t}_i (i = 1, 2)$ and the positive sign for top quark $t$,
and $n_l$ ($l$ = $t$, ${\tilde t}_2$, ${\tilde t}_2$) are the degrees of freedom
for top and stop quarks counting their charge, color, and spin factors.
Specifically, $n_t = -12$ for top quark and $n_{{\tilde t}_i} = 6$ $(i = 1, 2)$ for stop quarks.
The above radiative corrections should be added to the potential at the one-loop level
at zero temperature to yield the finite-temperature one-loop effective potential.
The above potential may have several minima, of which some may be degenerate.
The global minima of the above potential is defined as the vacua at $T$.

Let us denote the VEVs of the neutral Higgs fields at $T$ as $v_1(T)$, $v_2(T)$, and $x(T)$.
Hereafter, we will omit the temperature dependence of these VEVs,
except the zero-temperature values, namely, $v_1(0)$, $v_2(0)$, and $x(0)$.
In terms of the VEVs at $T$, the finite-temperature one-loop effective potential
may be written as
\begin{equation}
\langle V(v_1, v_2, x,T) \rangle = \langle V_0 \rangle + \langle V_1 \rangle + \langle V_T \rangle \ .
\end{equation}
Explicitly, we have
\begin{eqnarray}
\langle V_0 \rangle & = & {g_1^2 + g_2^2 \over 8} (v_1^2 - v_2^2)^2
+ \lambda^2 (v_1^2 v_2^2 + v_1^2 x^2 + v_2^2 x^2)  + m_1^2 v_1^2 + m_2^2 v_2^2 + m_3^2 x^2 \cr
& &\mbox{} - 2 \lambda A_{\lambda} v_1 v_2 x \cos \phi_0 - 2 \xi^3 x  \ , \cr
\langle V_1 \rangle & = & \sum_{i = 1}^2 {3 m_{{\tilde t}_i}^4 \over 32 \pi^2}
  \left (\log {m_{{\tilde t}_i}^2 \over \Lambda^2} - {3\over 2} \right )
  - {3 m_t^4 \over 16 \pi^2} \left (\log { m_t^2 \over \Lambda^2}
  - {3\over 2} \right ) \ , \cr
\langle V_T \rangle & = & \mbox{} - {6 T^4 \over \pi^2}
   \int_0^{\infty} dx \ x^2 \
   \log \left [1 + \exp{\left ( - \sqrt {x^2+{m_t^2(v_2)/T^2 }} \right )  } \right ] \cr
& &\mbox{} + \sum_{i=1}^2 {3 T^4 \over \pi^2}
   \int_0^{\infty} dx \ x^2 \
   \log \left [1 - \exp{\left ( - \sqrt {x^2+ m_{{\tilde t}_i}^2(v_1, v_2, x)/T^2 } \right )  } \right ]
\end{eqnarray}
where $\phi_0$ is a redefined phase from $\theta_1$ at the one-loop level
that arises from the CP odd tadpole minimum conditions, and
the soft SUSY breaking masses at the one-loop level are given by
\begin{eqnarray}
m_1^2 & = &\mbox{} - {m_Z^2 \over 2} \cos 2 \beta - \lambda^2 (x(0)^2 + v(0)^2 \sin^2 \beta)
                  + \lambda A_{\lambda} x(0) \tan \beta \cos \phi_0  \cr
& &\mbox{} + {3 h_t^2 \over 16 \pi^2} (\lambda^2 x^2 + \lambda A_t x \cot \beta \cos \phi_t)
 f(m_{{\tilde t}_1}^2, m_{{\tilde t}_2}^2)   \ , \cr
m_2^2 & = & {m_Z^2 \over 2} \cos 2 \beta - \lambda^2 (x(0)^2 + v(0)^2 \cos^2 \beta)
                  + \lambda A_{\lambda} x(0) \cot \beta \cos \phi_0  \cr
& &\mbox{} + {3 h_t^2 \over 16 \pi^2} (A_t^2 + \lambda A_t x \tan \beta \cos \phi_t)
 f(m_{{\tilde t}_1}^2, m_{{\tilde t}_2}^2)
- { 3 h_t^2 m_T^2 \over 16 \pi^2}
\log \left ( {m_{{\tilde t}_1}^2 m_{{\tilde t}_2}^2 \over \Lambda^4 e^2 } \right ) \cr
& &\mbox{} - { 3 h_t^2 m_t^2 \over 16 \pi^2}
\log \left ( {m_{{\tilde t}_1}^2 m_{{\tilde t}_2}^2 \over m_t^4} \right )
- { 3 h_t^2 \over 32 \pi^2} (m_{{\tilde t}_2}^2 - m_{{\tilde t}_1}^2)
\log  \left ({m_{{\tilde t}_2}^2  \over m_{{\tilde t}_1}^2} \right ) \ , \cr
m_3^2 & = &\mbox{} - \lambda^2 v(0)^2 + {\lambda \over 2 x(0)} v(0)^2 A_{\lambda} \sin 2 \beta \cos \phi_0
+ {\xi^3 \over x}  \cr
& &\mbox{} + {3 h_t^2 \lambda v(0)^2 \cot \beta \over 16 \pi^2 x(0)} (\lambda x(0) \cos \beta
+ A_t \sin \beta \cos \phi_t)
 f(m_{{\tilde t}_1}^2, m_{{\tilde t}_2}^2)  \ .
\end{eqnarray}
As mentioned before, $v_1(0)$, $v_2(0)$ and $x(0)$ are the vacuum expectation values evaluated
at zero temperature.

Notice that there is a trilinear term with $A_{\lambda}$ in $\langle V_0 \rangle$,
which will be shown to enhance the strength of the first-order EWPT.
Another trilinear term may also be found in $\langle V_T \rangle$,
arising from the thermal effects due to the stop quarks, which can be seen
by expanding the thermal potential using for example the high-temperature approximation.
However, we would not use the high-temperature approximation
but perform the exact numerical integration in the thermal potential.

Now, we inspect the above finite-temperature one-loop effective potential
in order to study at what temperature it may have two degenerate vacua.
This temperature is important for the EWPT, because it is the critical temperature, $T_c$,
at which the EWPT may take place from the symmetric phase to the broken phase.
Since vacua are the minima of the above potential,
the first derivatives of the potential with respect to the VEVs should vanish,
that is, $\partial V/\partial \varphi|_{\varphi = \langle \varphi \rangle} = 0$.

The first derivative of the above potential with respect to $x$ yields a minimum condition,
which may be written as
\begin{eqnarray}
    0 & = & 2 m_3^2 x - 2 \lambda A_{\lambda} v_1 v_2 \cos \phi_0 + 2 \lambda^2 (v_1^2 + v_2^2) x
    - 2 \xi^3 \cr
& & \mbox{} - {3 h_t^2 \lambda v_1 \over 8 \pi^2} (\lambda v_1 x + A_t v_2 \cos \phi_t)
 f(m_{{\tilde t}_1}^2, m_{{\tilde t}_2}^2) \cr
& & \mbox{} - {3 T^2 \over 2 \pi^2} {2 h_t^2 \lambda v_1 \over (m_{{\tilde t}_2}^2 - m_{{\tilde t}_1}^2) }
(\lambda x v_1 + A_t v_2 \cos \phi_t ) \cr
& &\mbox{} \times   \int_0^{\infty} dx \ x^2 \
    {\exp (-\sqrt{x^2 + m_{{\tilde t}_1}^2/T^2 })  \over \sqrt{x^2 + m_{{\tilde t}_1}^2/T^2 }
\{1 - \exp (-\sqrt{x^2 + m_{{\tilde t}_1}^2/T^2 }) \}}  \cr
& &\mbox{} + {3 T^2 \over 2 \pi^2} {2 h_t^2 \lambda v_1 \over (m_{{\tilde t}_2}^2 - m_{{\tilde t}_1}^2) }
(\lambda x v_1 + A_t v_2 \cos \phi_t )
   \int_0^{\infty} dx \ x^2  \cr
& &\mbox{} \times { \exp (-\sqrt{x^2 + m_{{\tilde t}_2}^2/T^2 })
\over \sqrt{x^2 + m_{{\tilde t}_2}^2/T^2 }
\{1 - \exp (-\sqrt{x^2 + m_{{\tilde t}_2}^2/T^2 }) \}}   \ .
\end{eqnarray}

In principle, this minimum condition can be used to express $x$ in terms of other variables.
In practice, however, the minimum condition cannot be solved analytically, since it is nonlinear.
We use the bisection method to solve it and obtain $x$.
Substituting $x$ into $\langle V(v_1,v_2,x,T) \rangle$, we can eliminate $x$
and obtain numerically the values of $\langle V(v_1,v_2, T) \rangle$.

By varying the temperature and searching over the $(v_1, v_2)$-plane,
we investigate $\langle V(v_1,v_2, T) \rangle$ in order to find out the critical temperature at which
it has two degenerate vacua.
Let us denote the two degenerate vacua, at the critical temperature,
as ($v_{1A},v_{2A}, x_A$) and ($v_{1B},v_{2B}, x_B$).
The EWPT may take place from the vacuum with symmetric phase to the vacuum with broken phase.
Let ($v_{1A},v_{2A}, x_A$) be the symmetric-phase state and ($v_{1B},v_{2B}, x_B$) be the broken-phase state.
The distance between the two vacua, at the critical temperature, is given by
\begin{equation}
v_c = \sqrt{(v_{1B}-v_{1A})^2 +(v_{2B}-v_{2A})^2 + (x_B-x_A)^2} \ ,
\end{equation}
and it is called as the critical VEV [31].

The strength of the first-order EWPT, which is driven by a non-equilibrium (out-of-equilibrium) condition,
depends on the relative size between the critical temperature $T_c$
and the critical VEV $v_c$.
In general, the first-order EWPT is said to be strong if $v_c>T_c$.
The main aim of our numerical analysis in the next section is therefore
to establish the possibility of a strongly first-order EWPT in the MNMSSM
with explicit CP violation by examining $v_c$ and $T_c$.

\section{NUMERICAL ANALYSIS}

For our numerical analysis, we fix $\Lambda = 300$ GeV for the renormalization scale
and $m_t = 175$ GeV for top quark mass.
The CP symmetry in the MNMSSM at the one-loop level at zero temperature is
explicitly violated by $\phi_t$, arising from the stop quark contributions
in the effective potential at zero temperature, which is in fact the only source of CP violation.
Other relevant free parameters besides $\phi_t$ in the MNMSSM are: $\tan \beta$,
$\lambda$, $A_{\lambda}$, $x(0)$, $\xi$, $m_T$, $A_t$, and $T$.

At given temperature, we examine the value of $\langle V(v_1, v_2, T)\rangle$
in the $(v_1, v_2)$-plane, where $x$ has been eliminated numerically by means of the minimum condition.
This job is repeated by varying the temperature.
As an illustration, we plot the equipotential contours of $\langle V(v_1, v_2, T)\rangle$
in the ($v_1,v_2$)-plane in Fig. 1 for $T=156$ GeV.
The remaining parameters are set as $\phi_t=\pi/100$, $\tan \beta=10$, $\lambda=0.3$,
$A_{\lambda}=500$ GeV, $x(0)=50$ GeV, $\xi= 80$ GeV, $m_T = 500$ GeV, and $A_t = 100$ GeV.

In Fig. 1, one can see that there are three stationary points for $\langle V(v_1, v_2, T)\rangle$.
The values of $\langle V(v_1, v_2, T)\rangle$ at these stationary points tell us that the point
at around $(v_1, v_2) = (170,320)$ GeV is a saddle point whereas the other two,
at around $(v_1, v_2) = (6,82)$ GeV and $(343,472)$ GeV, are minima.
Moreover, the values of $\langle V(v_1, v_2, T)\rangle$ at these two minima are found to be equal.
Therefore, they are degenerate minima, with a potential barrier between them.
The first-order EWPT may take place from one vacuum to the other one through thermal tunnelling.

The shape of $\langle V(v_1, v_2, T)\rangle$ in Fig. 1 is actually a typical example for the first-order EWPT.
The temperature of $T = 156$ GeV is the critical temperature $T_c$
that allows the phase transition from the symmetric-phase state to the broken-phase state.
It is found that the minimum point at around $(v_1, v_2) = (6,82)$ GeV is
the symmetric-phase vacuum whereas the minimum point at around $(v_1, v_2) = (343,472)$ GeV is
the broken-phase vacuum.

In order to estimate the strength of the first-order EWPT for the potential in Fig. 1,
we need to know $v_c$.
We obtain $x = 34$ GeV and $x =519$ GeV respectively for $(v_1, v_2) = (6,82)$ GeV
and $(v_1, v_2) = (343,472)$ GeV.
Thus, the distance between $(v_{1A},v_{2A},x_A)=(6,82,34)$ GeV and $(v_{1B},v_{2B},x_B)=(343,472,519)$ GeV,
where the subscripts $A$ and $B$ denote respectively the symmetric-phase state and the broken-phase one,
is roughly $v_c = 708$ GeV.
Comparing this value with the critical temperature, $T_c = 156$ GeV, we obtain $v_c/T_c \sim 4.53$.
This ratio implies that the first-order EWPT shown in Fig. 1 is strong.

The numerical analysis of Fig. 1 suggests that there is at least one set of free parameters of the MNMSSM
with explicit CP violation at the one-loop level which allows the possibility of a strongly first-order EWPT.
For the parameter set of Fig. 1, other relevant values are evaluated.
The masses of the neutral Higgs boson are calculated to be $m_{h_1} = 60.91$ GeV, $m_{h_2} = 94.92$ GeV,
$m_{h_3} = 130.27$ GeV, $m_{h_4} = 292.75$ GeV, and $m_{h_5} = 296.03$ GeV.
The effective size of the CP violation is $\rho = 0.47 \times 10^{-4}$, which is very small,
since $\phi_t = \pi/100$.
The masses of the stop quarks are calculated to be $m_{{\tilde t}_1} = 512.70$ GeV
and $m_{{\tilde t}_2} = 546.24$ GeV.
They are much heavier than the top quark.

By examining other sets of parameters, we find that a strongly first-order EWPT is possible for
a wide parameter space of the MNMSSM with explicit CP violation at the one-loop level.
After establishing the possibility of the strongly first-order EWPT,
we are now interested in the dependence of the strength of the first-order EWPT on such factors
as the CP phase $\phi_t$, the mass of the lightest Higgs boson, and the stop quark masses.

First, in order to study the effect of the CP phase on the strength of the first-order EWPT,
we take $\phi_t=\pi/2$, which is maximal since $\sin\phi_t = 1$.
The other parameters are set for simplicity as Fig. 1: $\tan \beta=10$, $\lambda=0.3$,
$A_{\lambda}=500$ GeV, $x(0)=50$ GeV, $\xi= 80$ GeV, $m_T = 500$ GeV, and $A_t =100$ GeV.
At $T_c = 161$ GeV, we find that the potential has three stationary points with required properties.
The result is shown in Fig. 2, where the equipotential contours of $\langle V \rangle$ are plotted
in the ($v_1, v_2$)-plane at that temperature.
The coordinates in the ($v_1, v_2$)-plane of the two degenerate minima of $\langle V \rangle$
are $(v_{1A}, v_{2A})=(6,52)$ GeV with $x_A = 33$ GeV and $(v_{1B}, v_{2B})=(356,487)$ GeV with $x_B = 533$ GeV.
The strength of the first-order EWPT is $v_c/T_c= 4.65$.
We also have $\rho = 0.76 \times 10^{-3}$.
For the masses of relevant particles, we obtain $m_{h_1} = 60.60$ GeV, $m_{h_2} = 94.92$ GeV,
$m_{h_3} = 130.39$ GeV, $m_{h_4} = 293.38$ GeV, and $m_{h_5} = 296.66$ GeV for neutral Higgs bosons,
and $m_{{\tilde t}_1} = 512.95$ GeV and $m_{{\tilde t}_2} = 546.01$ GeV for the stop quarks.

The numerical results in Fig. 2 are essentially the same as those in Fig. 1, although $\phi_t$ is
respectively nearly zero in Fig. 1 and maximal in Fig. 2.
The mass spectra of neutral Higgs bosons or the stop quarks are almost the same;
the critical temperature obtained in Fig. 2 is slightly higher than that in Fig. 1;
and the strength of the first-order EWPT in Fig. 2 is also a little stronger than that in Fig. 1.
It is thus hard to draw any conclusive relationship between the effect of $\phi_t$
and the strength of the first-order EWPT, by comparing the numerical results of Fig. 2
with those in Fig. 1 alone.
We may suggest that, in the MNMSSM with explicit CP violation, the CP phase $\phi_t$ enhances slightly
the first-order EWPT.
This behavior of the present model might be compared with the results of other supersymmetric models.
In the NMSSM with $Z_3$ symmetry, for example, the CP phase $\phi_t$ reduces a little the strength of
the first-order EWPT [27].

One might suggest that the strength of the first-order EWPT is dependent on the lightest Higgs boson mass.
Note that the lightest Higgs boson mass in Fig. 2 is smaller, negligibly, than that in Fig. 1.
Thus, one can expect that the strength of the EWPT is enhanced as the lightest Higgs boson becomes light.
This is reasonable in both the SM and the MSSM, where the strength of the EWPT may always increase
when the lightest Higgs boson mass decreases.
Also, the non-minimal supersymmetric models with a Higgs singlet can accommodate a stronger EWPT
for a lighter $h_1$ [28,29].

In order to clarify whether or not the smaller lightest Higgs boson mass may indeed enhance
the strength of the first-order EWPT, in the present model, we study by changing the parameter values.
We take $\lambda=0.2$ and $\xi=100$ GeV in order to increase the lightest Higgs boson mass.
The other parameters are the same as Fig. 2.
That is, $\phi_t = \pi/2$, $\tan \beta = 10$, $A_{\lambda}=500$ GeV, $x(0) = 50$ GeV, $m_T = 500$ GeV,
and $A_t = 100$ GeV.
Note in particular that the CP phase $\phi_t=\pi/2$ is the same as Fig. 2.

At  $T_c =191$ GeV, we obtain Fig. 3.
The potential has two degenerate minima at $(v_{1A},v_{2A},x_A) = (11,11,40)$ GeV
and $(v_{1B},v_{2B},x_B) =(686,753,784)$ GeV.
The strength of the first-order EWPT is about $v_c/T_c = 6.53$.
Also, we have $\rho = 0.17 \times 10^{-2}$.
The mass spectra of relevant particles are: $m_{h_1} = 89.96$ GeV, $m_{h_2} = 130.15$ GeV,
$m_{h_3} = 144.88$ GeV, $m_{h_4} = 244.18$ GeV, and $m_{h_5} = 247.14$ GeV for the neutral Higgs bosons,
and $m_{{\tilde t}_1} = 512.95$ GeV and $m_{{\tilde t}_2} = 546.0$ GeV for the stop quarks.

Comparing Fig. 3 with Fig. 2, where $m_{h_1} = 89.96$ GeV with $v_c/T_c = 6.53$ in Fig. 3
while $m_{h_1} = 60.60$ GeV with $v_c/T_c= 4.65$ in Fig. 2, one might argue
that the strength of the EWPT is increases as the lightest Higgs boson mass increases.
On the other hand, by comparing Fig. 2 with Fig. 1, one might obtain an opposite argument.
Thus, it is likely that the strength of the EWPT in the MNMSSM with explicit CP violation
does not consistently dependent on the variation of the lightest Higgs boson mass.
It can be either enhanced or reduced as the lightest Higgs boson mass increases.

Now, we investigate the dependence of the strength of the EWPT on the stop quark masses.
It has been discussed in the literature that the stop quarks play some crucial role
in other supersymmetric models within the context of the EWPT.
The MSSM always need a light stop quark below top quark mass in order to achieve
a strongly first-order EWPT, whether or not CP violation occurs in its Higgs sector.
This behavior of the MSSM is understandable,
since the thermal contribution of the stop quarks is suppressed as they become very heavy.
On the other hand, the MNMSSM with explicit CP violation is known to have a parameter set,
without need of a light stop quark, where the first-order EWPT
is as much strong as in the case of Fig. 1 in the present model.

In the present model, there is the trilinear term containing $A_{\lambda}$,
already at the tree level, which leads to a stronger EWPT.
As seen in the previous figures, a strongly first-order EWPT is allowed
without a very light stop quark in the MNMSSM with explicit CP violation.
We take $m_T=600$ GeV in order to increase the stop quark masses.
The other parameters are the same as Fig. 3: $\phi_t = \pi/2$, $\tan \beta = 10$,
$\lambda= 0.2$, $A_{\lambda} = 500$ GeV, $x(0) = 50$ GeV, $\xi = 100$ GeV, and $A_t = 100$ GeV.
With these parameter values, Fig. 4 shows for $T_c=145$
the equipotential contours of $\langle V \rangle$ in the ($v_1, v_2$)-plane
with two degenerate vacua at $(v_{1A},v_{2A},x_A) = (11,123,44)$ GeV
and $(v_{1B},v_{2B},x_B) =  (641,708,748)$ GeV.
The strength of the first-order EWPT is about $v_c/T_c = 7.66$.
We obtain $\rho = 0.13 \times 10^{-2}$, and $m_{h_1} = 92.66$ GeV,
$m_{h_2} = 130.15$ GeV, $m_{h_3} = 145.68$ GeV, $m_{h_4} = 244.18$ GeV,
and $m_{h_5} = 247.20$ GeV for the neutral Higgs boson masses
and $m_{{\tilde t}_1} = 610.838$ GeV and $m_{{\tilde t}_2} = 638.84$ GeV
for the stop quark masses.

These stop quark masses may be compared with those in Fig. 3, namely,
$m_{{\tilde t}_1} = 512.95$ GeV and $m_{{\tilde t}_2} = 546.0$ GeV.
The stop quark masses in Fig. 4 are increased but not so significantly changed from Fig. 3.
Nevertheless, we find that the strength of the first order EWPT is noticeably increased.
Thus, we may deduce that heavy stop quarks are favored by the strongly first-order EWPT,
in the MNMSSM with explicit CP violation.
This behavior distinguishes the present model from the MSSM.

\section{CONCLUSIONS}

We have established the possibility of a strongly first-order EWPT in the MNMSSM
with explicit CP violation.
At the tree level, the Higgs potential of the model can be made free from any complex CP phase
by suitably rotating the Higgs fields.
At the one-loop level, as the radiative corrections due to the stop quarks
with non-degenerate masses are taken into account, the explicit CP violation can be generated
by a CP phase $\phi_t$.
We have studied the Higgs sector of the model by considering the finite-temperature
one-loop effective potential, where the thermal effects due to top and stop quarks
are numerically evaluated by exact integration.
We find that some parameter space of the model allows a strongly first-order EWPT.

We also have studied how a strongly first-order EWPT depends on the effective size of CP violation,
the neutral Higgs boson masses, and the stop quark masses for given sets of parameter values.
We find that the strength of the EWPT has not a consistent dependence on
the variation of the mass of the lightest Higgs boson:
It can be either enhanced or reduced with increasing the lightest Higgs boson mass.
We find that the strength of the EWPT increases as the complex CP phase gets large.
It also becomes stronger as the stop quarks get heavier.

\vskip 0.3 in
\noindent
{\large {\bf ACKNOWLEDGMENTS}}
\vskip 0.2 in
\noindent
This research is supported by KOSEF through CHEP.
The authors would like to acknowledge the support from KISTI
(Korea Institute of Science and Technology Information) under
"The Strategic Supercomputing Support Program" with Dr. Kihyeon Cho as the technical supporter.
The use of the computing system of the Supercomputing Center is also appreciated.

\vfil\eject



\vfil\eject
{\noindent\bf Figure Captions}

\vskip 0.2 in
\noindent
Fig. 1 : Equipotential contours of $\langle V \rangle$ at $T_c=156$ GeV
in the ($v_1, v_2$)-plane for $\phi_t = \pi/100$,
$\tan \beta = 10$, $\lambda= 0.3$, $A_{\lambda} = 500$ GeV, $x(0) = 50$ GeV, $\xi = 80$ GeV,
$m_T = 500$ GeV, and $A_t = 100$ GeV.
The symmetric-phase vacuum is located at $(v_1,v_2) = (6,82)$ GeV with $x = 34$ GeV
and the broken-phase vacuum is located at  $(v_1,v_2) = (343,472)$ GeV with $x = 519$ GeV.
The strength of the first order EWPT is about $v_c/T_c = 4.53$.
We have $\rho = 0.47 \times 10^{-4}$, and $m_{h_1} = 60.91$ GeV, $m_{h_2} = 94.92$ GeV,
$m_{h_3} = 130.27$ GeV,
$m_{h_4} = 292.75$ GeV, and $m_{h_5} = 296.03$ GeV for the neutral Higgs boson masses
and  $m_{{\tilde t}_1} = 512.70$ GeV and $m_{{\tilde t}_2} = 546.24$ GeV for the stop quark masses.

\vskip 0.15 in
\noindent
Fig. 2 : Equipotential contours of $\langle V \rangle$ at $T_c=135$ GeV
in the ($v_1, v_2$)-plane for $\phi_t = \pi/2$,
$\tan \beta = 10$, $\lambda= 0.3$, $A_{\lambda} = 500$ GeV, $x(0) = 50$ GeV, $\xi = 80$ GeV,
$m_T = 500$ GeV, and $A_t = 100$ GeV.
The symmetric-phase vacuum is located at $(v_1,v_2) = (6,52)$ GeV with $x = 33$ GeV
and the broken-phase vacuum is located at  $(v_1,v_2) = (356,487)$ GeV with $x = 533$ GeV.
The strength of the first order EWPT is about $v_c/T_c = 4.65$.
We have $\rho =  0.76 \times 10^{-3}$, and $m_{h_1} = 60.60$ GeV, $m_{h_2} = 94.92$ GeV,
$m_{h_3} = 130.39$ GeV,
$m_{h_4} = 293.38$ GeV, and $m_{h_5} = 296.66$ GeV for the neutral Higgs boson masses
and  $m_{{\tilde t}_1} = 512.95$ GeV and $m_{{\tilde t}_2} = 546.01$ GeV for the stop quark masses.

\vskip 0.15 in
\noindent
Fig. 3 : Equipotential contours of $\langle V \rangle$ at $T_c=191$ GeV
in the ($v_1, v_2$)-plane for $\phi_t = \pi/2$,
$\tan \beta = 10$, $\lambda= 0.2$, $A_{\lambda} = 500$ GeV, $x(0) = 50$ GeV, $\xi = 100$ GeV,
$m_T = 500$ GeV, and $A_t = 100$ GeV.
The symmetric-phase vacuum is located at $(v_1,v_2) = (11,11)$ GeV
with $x = 40$ GeV and the broken-phase vacuum is located at  $(v_1,v_2) = (686,753)$ GeV with $x = 784$ GeV.
The strength of the first order EWPT is about $v_c/T_c = 6.53$.
We have $\rho =  0.17 \times 10^{-2}$, and $m_{h_1} = 89.96$ GeV, $m_{h_2} = 130.15$ GeV,
$m_{h_3} = 144.88$ GeV, $m_{h_4} = 244.18$ GeV, and $m_{h_5} = 247.14$ GeV
for the neutral Higgs boson masses and  $m_{{\tilde t}_1} = 512.95$ GeV
and $m_{{\tilde t}_2} = 546.0$ GeV for the stop quark masses.

\vskip 0.15 in
\noindent
Fig. 4 : Equipotential contours of $\langle V \rangle$ at $T_c=145$ GeV
in the ($v_1, v_2$)-plane for $\phi_t = \pi/2$,
$\tan \beta = 10$, $\lambda= 0.2$, $A_{\lambda} = 500$ GeV, $x(0) = 50$ GeV, $\xi = 100$ GeV,
$m_T = 600$ GeV, and $A_t = 100$ GeV.
The symmetric-phase vacuum is located at $(v_1,v_2) = (11,123)$ GeV
with $x = 44$ GeV and the broken-phase vacuum is located at  $(v_1,v_2) = (641,708)$ GeV with $x = 748$ GeV.
The strength of the first order EWPT is about $v_c/T_c = 7.66$.
We have $\rho =  0.13 \times 10^{-2}$, and $m_{h_1} = 92.66$ GeV,
$m_{h_2} = 130.15$ GeV, $m_{h_3} = 145.68$ GeV, $m_{h_4} = 244.18$ GeV,
and $m_{h_5} = 247.20$ GeV for the neutral Higgs boson masses and
$m_{{\tilde t}_1} = 610.83$ GeV and $m_{{\tilde t}_2} = 638.84$ GeV for the stop quark masses.

\vfil\eject

\renewcommand\thefigure{1}
\begin{figure}[t]
\begin{center}
\includegraphics[scale=0.6]{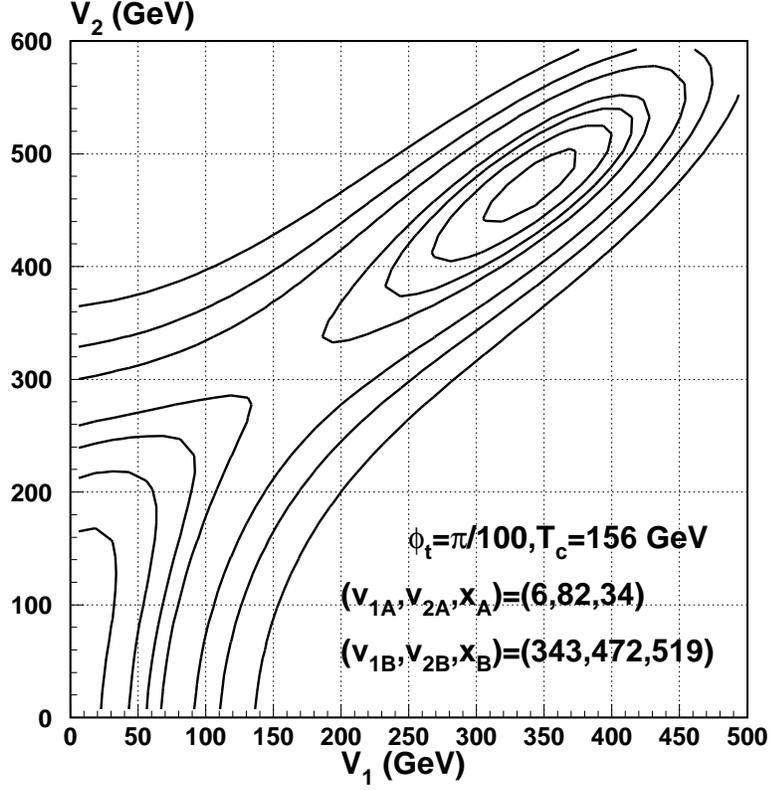}
\caption[plot]{Equipotential contours of $\langle V \rangle$ at $T_c=156$ GeV
in the ($v_1, v_2$)-plane for $\phi_t = \pi/100$, $\tan \beta = 10$, $\lambda= 0.3$,
$A_{\lambda} = 500$ GeV, $x(0) = 50$ GeV, $\xi = 80$ GeV,
$m_T = 500$ GeV, and $A_t = 100$ GeV.
The symmetric-phase vacuum is located at $(v_1,v_2) = (6,82)$ GeV with $x = 34$ GeV
and the broken-phase vacuum is located at  $(v_1,v_2) = (343,472)$ GeV with $x = 519$ GeV.
The strength of the first order EWPT is about $v_c/T_c = 4.53$.
We have $\rho = 0.47 \times 10^{-4}$, and $m_{h_1} = 60.91$ GeV, $m_{h_2} = 94.92$ GeV, $m_{h_3} = 130.27$ GeV,
$m_{h_4} = 292.75$ GeV, and $m_{h_5} = 296.03$ GeV for the neutral Higgs boson masses
and  $m_{{\tilde t}_1} = 512.70$ GeV and $m_{{\tilde t}_2} = 546.24$ GeV for the stop quark masses.}
\end{center}
\end{figure}

\renewcommand\thefigure{2}
\begin{figure}[t]
\begin{center}
\includegraphics[scale=0.6]{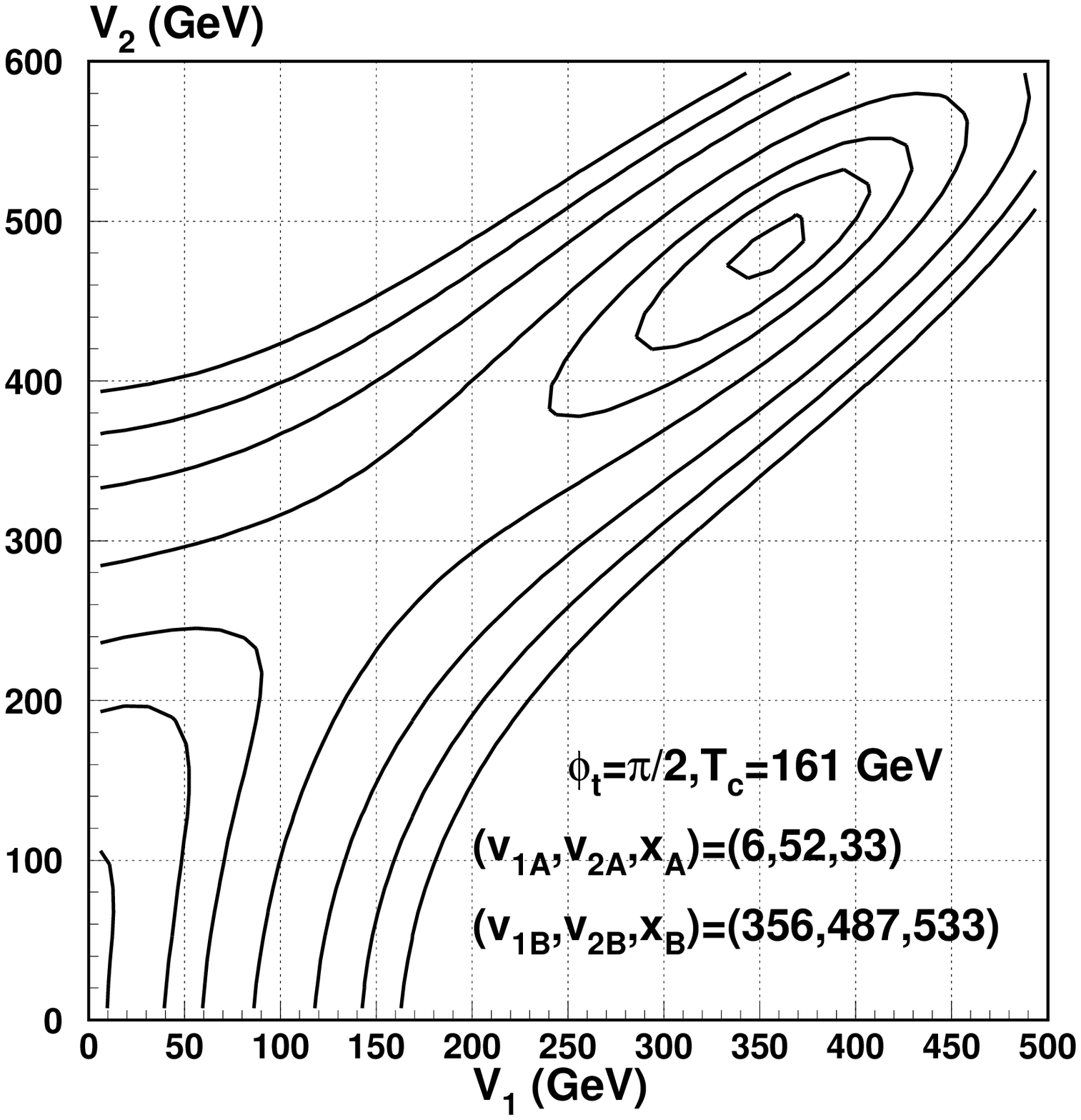}
\caption[plot]{Equipotential contours of $\langle V \rangle$ at $T_c=135$ GeV
in the ($v_1, v_2$)-plane for $\phi_t = \pi/2$, $\tan \beta = 10$, $\lambda= 0.3$,
$A_{\lambda} = 500$ GeV, $x(0) = 50$ GeV, $\xi = 80$ GeV,
$m_T = 500$ GeV, and $A_t = 100$ GeV.
The symmetric-phase vacuum is located at $(v_1,v_2) = (6,52)$ GeV with $x = 33$ GeV
and the broken-phase vacuum is located at  $(v_1,v_2) = (356,487)$ GeV with $x = 533$ GeV.
The strength of the first order EWPT is about $v_c/T_c = 4.65$.
We have $\rho =  0.76 \times 10^{-3}$, and $m_{h_1} = 60.60$ GeV, $m_{h_2} = 94.92$ GeV,
$m_{h_3} = 130.39$ GeV,
$m_{h_4} = 293.38$ GeV, and $m_{h_5} = 296.66$ GeV for the neutral Higgs boson masses
and  $m_{{\tilde t}_1} = 512.95$ GeV and $m_{{\tilde t}_2} = 546.01$ GeV for the stop quark masses.}
\end{center}
\end{figure}

\renewcommand\thefigure{3}
\begin{figure}[t]
\begin{center}
\includegraphics[scale=0.6]{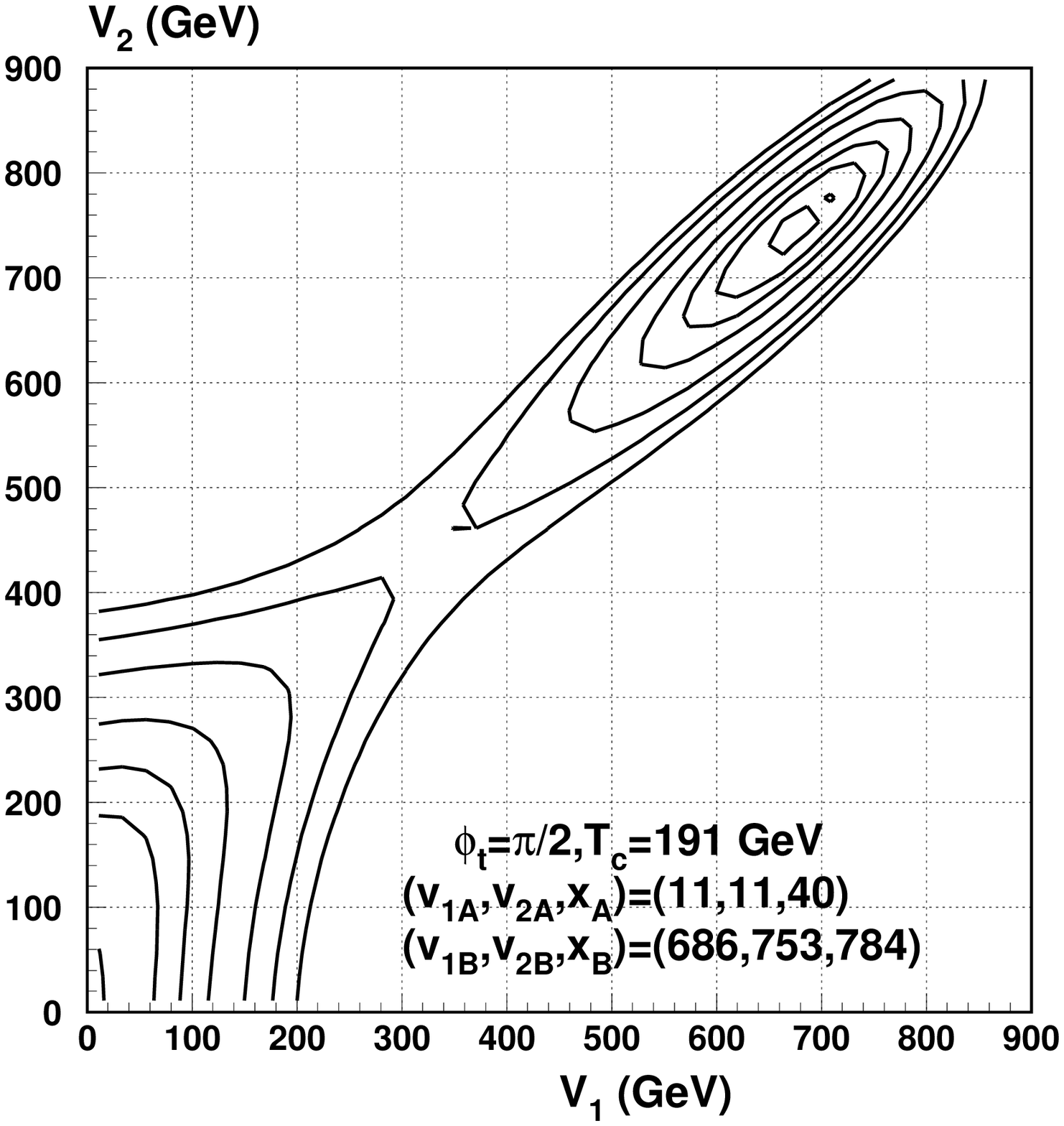}
\caption[plot]{Equipotential contours of $\langle V \rangle$ at $T_c=191$ GeV
in the ($v_1, v_2$)-plane for $\phi_t = \pi/2$, $\tan \beta = 10$, $\lambda= 0.2$,
$A_{\lambda} = 500$ GeV, $x(0) = 50$ GeV, $\xi = 100$ GeV,
$m_T = 500$ GeV, and $A_t = 100$ GeV.
The symmetric-phase vacuum is located at $(v_1,v_2) = (11,11)$ GeV with $x = 40$ GeV
and the broken-phase vacuum is located at  $(v_1,v_2) = (686,753)$ GeV with $x = 784$ GeV.
The strength of the first order EWPT is about $v_c/T_c = 6.53$.
We have $\rho =  0.17 \times 10^{-2}$, and $m_{h_1} = 89.96$ GeV, $m_{h_2} = 130.15$ GeV,
$m_{h_3} = 144.88$ GeV, $m_{h_4} = 244.18$ GeV, and $m_{h_5} = 247.14$ GeV
for the neutral Higgs boson masses and  $m_{{\tilde t}_1} = 512.95$ GeV
and $m_{{\tilde t}_2} = 546.00$ GeV for the stop quark masses.}
\end{center}
\end{figure}

\renewcommand\thefigure{4}
\begin{figure}[t]
\begin{center}
\includegraphics[scale=0.6]{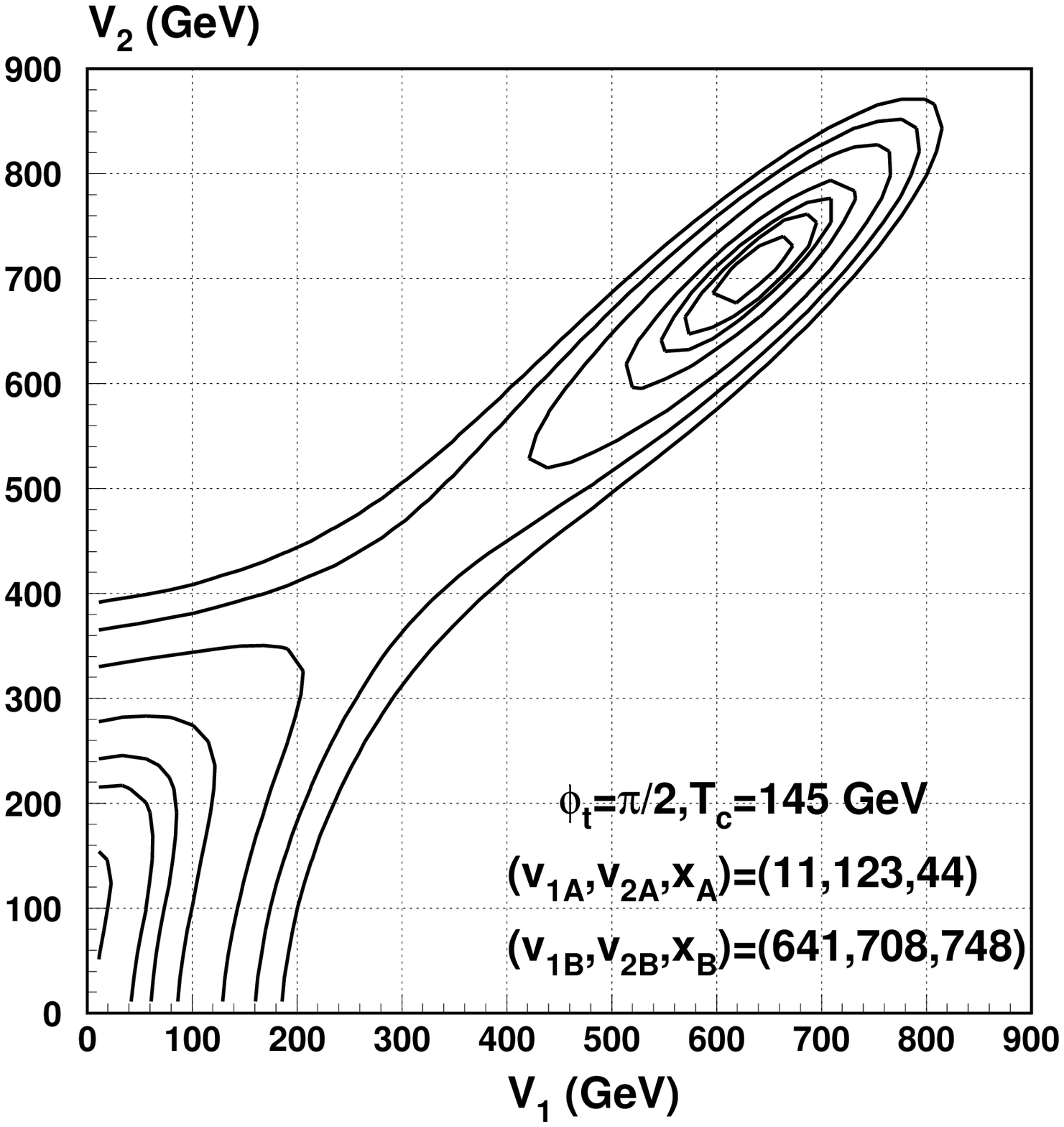}
\caption[plot]{Equipotential contours of $\langle V \rangle$ at $T_c=145$ GeV
in the ($v_1, v_2$)-plane for $\phi_t = \pi/2$, $\tan \beta = 10$, $\lambda= 0.2$,
$A_{\lambda} = 500$ GeV, $x(0) = 50$ GeV, $\xi = 100$ GeV,
$m_T = 600$ GeV, and $A_t = 100$ GeV.
The symmetric-phase vacuum is located at $(v_1,v_2) = (11,123)$ GeV
with $x = 44$ GeV and the broken-phase vacuum is located at  $(v_1,v_2) = (641,708)$ GeV
with $x = 748$ GeV.
The strength of the first order EWPT is about $v_c/T_c = 7.66$.
We have $\rho =  0.13 \times 10^{-2}$, and $m_{h_1} = 92.66$ GeV,
$m_{h_2} = 130.15$ GeV, $m_{h_3} = 145.68$ GeV, $m_{h_4} = 244.18$ GeV,
and $m_{h_5} = 247.20$ GeV for the neutral Higgs boson masses
and  $m_{{\tilde t}_1} = 610.83$ GeV and $m_{{\tilde t}_2} = 638.84$ GeV
for the stop quark masses.}
\end{center}
\end{figure}

\end{document}